\documentclass[a4paper,  10pt]{IEEEtran}  
\usepackage{fancyhdr}
\usepackage{makecell}
\usepackage{psfrag}
\usepackage{amsmath,amssymb,amsfonts}
\usepackage{siunitx} 
\usepackage[inkscapelatex=false]{svg}
                    
\usepackage{dsfont,latexsym,cite, comment, graphicx,color}
\usepackage{url, hyperref}
\usepackage[T1]{fontenc}

\usepackage{subfigure}

\usepackage[english]{babel}	

\title{\LARGE \bf Model Reference Control for Wind Turbine Systems in Full Load Region based on Takagi-Sugeno Fuzzy Systems}

\usepackage{tikz}
\usetikzlibrary{shapes,arrows,shadows,decorations.pathmorphing, decorations.pathreplacing, decorations.shapes,calc}
\usetikzlibrary{shadows} 

\tikzstyle{decision} = [diamond, draw, fill=yellow!10, 
    text width=4.5em, text badly centered, node distance=3cm, inner sep=0pt]
\tikzstyle{block} = [rectangle, draw, fill=yellow!10, 
    text width=25em, text centered, rounded corners, minimum height=4em]

\tikzstyle{cloud} = [draw, ellipse,fill=red!20, node distance=3cm, minimum height=2em]

\tikzstyle{arrow} = [draw, -latex']
\tikzstyle{line} = [draw]

\usepackage{varwidth}

\usepackage{booktabs}
\begin{document}


\author{
	\IEEEauthorblockN{Johannes Brunner, Jens Fortmann, Horst Schulte}\\
	\IEEEauthorblockA{University of Applied Sciences Berlin (HTW) \\
		Faculty: School of Engineering - Energy and Information Science\\
		Control Engineering Group, 12459 Berlin, Germany\\
		Email: \{brunner, schulte\}@htw-berlin.de}
}



\newcommand{\stylematrix}[1]{\boldsymbol{\mathrm{#1}}}
\newcommand{\styleset}[1]{\mathit{#1}}
\newcommand{\stylescalar}[1]{\mathit{#1}}
\newcommand{\stylevector}[1]{\boldsymbol{#1}}
\newcommand{\substyle}[1]{\mathrm{#1}}
\newcommand{\discr}{^\text{d}}


\newcommand{\ct}{\ensuremath{\stylescalar{t}}}
\newcommand{\cn}{\ensuremath{\stylescalar{n}}}

\newcommand{\cA}{\ensuremath{\stylematrix{A}}}	
\newcommand{\cB}{\ensuremath{\stylematrix{B}}}	
\newcommand{\cC}{\ensuremath{\stylematrix{C}}}	
\newcommand{\cD}{\ensuremath{\stylematrix{D}}}	
\newcommand{\cG}{\ensuremath{\stylematrix{G}}}
\newcommand{\cP}{\ensuremath{\stylematrix{P}}}
\newcommand{\cM}{\ensuremath{\stylematrix{M}}}
\newcommand{\cX}{\ensuremath{\stylematrix{X}}}
\newcommand{\cS}{\ensuremath{\stylematrix{S}}}
\newcommand{\cK}{\ensuremath{\stylematrix{K}}}
\newcommand{\cQ}{\ensuremath{\stylematrix{Q}}}
\newcommand{\cI}{\ensuremath{\stylematrix{I}}}
\newcommand{\cR}{\ensuremath{\stylematrix{R}}}
\newcommand{\cL}{\ensuremath{\stylematrix{L}}}
\newcommand{\cW}{\ensuremath{\stylematrix{W}}}
\newcommand{\cY}{\ensuremath{\stylematrix{Y}}}

\newcommand{\cAd}{\ensuremath{\stylematrix{A}\discr}}	
\newcommand{\cBd}{\ensuremath{\stylematrix{B}\discr}}	
\newcommand{\cCd}{\ensuremath{\stylematrix{C}\discr}}	
\newcommand{\cDd}{\ensuremath{\stylematrix{D}\discr}}	
\newcommand{\cGd}{\ensuremath{\stylematrix{G}\discr}}
\newcommand{\cPd}{\ensuremath{\stylematrix{P}\discr}}
\newcommand{\cMd}{\ensuremath{\stylematrix{M}\discr}}
\newcommand{\cXd}{\ensuremath{\stylematrix{X}\discr}}
\newcommand{\cSd}{\ensuremath{\stylematrix{S}\discr}}
\newcommand{\cKd}{\ensuremath{\stylematrix{K}\discr}}
\newcommand{\cQd}{\ensuremath{\stylematrix{Q}\discr}}
\newcommand{\cId}{\ensuremath{\stylematrix{I}\discr}}
\newcommand{\cRd}{\ensuremath{\stylematrix{R}\discr}}
\newcommand{\cLd}{\ensuremath{\stylematrix{L}\discr}}
\newcommand{\cWd}{\ensuremath{\stylematrix{W}\discr}}
\newcommand{\cYd}{\ensuremath{\stylematrix{Y}\discr}}

\newcommand{\cAi}{\ensuremath{\stylematrix{A}_i}}	
\newcommand{\cBi}{\ensuremath{\stylematrix{B}_i}}	
\newcommand{\cCi}{\ensuremath{\stylematrix{C}_i}}	
\newcommand{\cDi}{\ensuremath{\stylematrix{D}_i}}
\newcommand{\cEi}{\ensuremath{\stylematrix{E}_i}}
\newcommand{\cFi}{\ensuremath{\stylematrix{F}_i}}
\newcommand{\cKi}{\ensuremath{\stylematrix{K}_i}}
\newcommand{\cMi}{\ensuremath{\stylematrix{M}_i}}
\newcommand{\cPi}{\ensuremath{\stylematrix{P}_i}}
\newcommand{\cXi}{\ensuremath{\stylematrix{X}_i}}

\newcommand{\cAj}{\ensuremath{\stylematrix{A}_j}}	
\newcommand{\cBj}{\ensuremath{\stylematrix{B}_j}}	
\newcommand{\cCj}{\ensuremath{\stylematrix{C}_j}}	
\newcommand{\cDj}{\ensuremath{\stylematrix{D}_j}}
\newcommand{\cEj}{\ensuremath{\stylematrix{E}_j}}
\newcommand{\cFj}{\ensuremath{\stylematrix{F}_j}}
\newcommand{\cKj}{\ensuremath{\stylematrix{K}_j}}
\newcommand{\cMj}{\ensuremath{\stylematrix{M}_j}}

\newcommand{\cGii}{\ensuremath{\stylematrix{G}_{ii}}}
\newcommand{\cGij}{\ensuremath{\stylematrix{G}_{ij}}}
\newcommand{\cGji}{\ensuremath{\stylematrix{G}_{ji}}}

\newcommand{\cbi}{\ensuremath{\stylevector{b}_{i}}}
\newcommand{\cai}{\ensuremath{\stylevector{a}_{i}}}
\newcommand{\cci}{\ensuremath{\stylevector{c}_{i}}}


\newcommand{\hij}{\ensuremath{\sum\limits_{i=1}^{N_r} \sum\limits_{j=1}^{N_r}h_i(\cz)h_j(\cz)}}
\newcommand{\hi}{\ensuremath{\sum\limits_{i=1}^{N_r}h_i(\cz)}}


\newcommand{\cAr}{\ensuremath{\stylematrix{A}^\text{r}}}	
\newcommand{\cBr}{\ensuremath{\stylematrix{B}}}	
\newcommand{\cCr}{\ensuremath{\stylematrix{C}^\text{r}}}	
\newcommand{\cDr}{\ensuremath{\stylematrix{D}^\text{r}}}
\newcommand{\cEr}{\ensuremath{\stylematrix{E}^\text{r}}}
\newcommand{\cFr}{\ensuremath{\stylematrix{F}^\text{r}}}

\newcommand{\cxr}{\ensuremath{\stylevector{x}^\text{r}}}
\newcommand{\cxI}{\ensuremath{\stylevector{x}_\text{I}}}
\newcommand{\cyr}{\ensuremath{\stylevector{y}^\text{r}}}

\newcommand{\cxbar}{\ensuremath{\Bar{\stylevector{x}}}}	

\newcommand{\cAibar}{\ensuremath{\Bar{\stylematrix{A}}_i}}
\newcommand{\cBibar}{\ensuremath{\Bar{\stylematrix{B}}_i}}	
\newcommand{\cCibar}{\ensuremath{\Bar{\stylematrix{C}}_i}}	
\newcommand{\cDibar}{\ensuremath{\Bar{\stylematrix{D}}_i}}
\newcommand{\cEibar}{\ensuremath{\Bar{\stylematrix{E}}_i}}
\newcommand{\cFibar}{\ensuremath{\Bar{\stylematrix{F}}_i}}
\newcommand{\cKibar}{\ensuremath{\Bar{\stylematrix{K}}_i}}
\newcommand{\cMibar}{\ensuremath{\Bar{\stylematrix{M}}_i}}
\newcommand{\cPibar}{\ensuremath{\Bar{\stylematrix{P}}_i}}
\newcommand{\cXibar}{\ensuremath{\Bar{\stylematrix{X}}_i}}

\newcommand{\cAjbar}{\ensuremath{\Bar{\stylematrix{A}}_j}}	
\newcommand{\cBjbar}{\ensuremath{\Bar{\stylematrix{B}}_j}}	
\newcommand{\cCjbar}{\ensuremath{\Bar{\stylematrix{C}}_j}}	
\newcommand{\cDjbar}{\ensuremath{\Bar{\stylematrix{D}}_j}}
\newcommand{\cEjbar}{\ensuremath{\Bar{\stylematrix{E}}_j}}
\newcommand{\cFjbar}{\ensuremath{\Bar{\stylematrix{F}}_j}}
\newcommand{\cKjbar}{\ensuremath{\Bar{\stylematrix{K}}_j}}
\newcommand{\cMjbar}{\ensuremath{\Bar{\stylematrix{M}}_j}}

\newcommand{\cGiibar}{\ensuremath{\Bar{\stylematrix{G}}_{ii}}}
\newcommand{\cGijbar}{\ensuremath{\Bar{\stylematrix{G}}_{ij}}}
\newcommand{\cGjibar}{\ensuremath{\Bar{\stylematrix{G}}_{ji}}}

\newcommand{\cbibar}{\ensuremath{\Bar{\stylevector{b}}_{i}}}
\newcommand{\caibar}{\ensuremath{\Bar{\stylevector{a}}_{i}}}
\newcommand{\ccibar}{\ensuremath{\Bar{\stylevector{c}}_{i}}}

\newcommand{\cx}{\ensuremath{\stylevector{x}}}	
\newcommand{\cu}{\ensuremath{\stylevector{u}}}	
\newcommand{\cf}{\ensuremath{\stylevector{f}}}	
\newcommand{\cz}{\ensuremath{\stylevector{z}}}
\newcommand{\cy}{\ensuremath{\stylevector{y}}}
\newcommand{\cg}{\ensuremath{\stylevector{g}}}	
\newcommand{\cv}{\ensuremath{\stylevector{v}}}	
\newcommand{\cd}{\ensuremath{\stylevector{d}}}	
\newcommand{\ce}{\ensuremath{\stylevector{e}}}	
\newcommand{\cs}{\ensuremath{\stylevector{s}}}
\newcommand{\cw}{\ensuremath{\stylevector{w}}}
\newcommand{\cc}{\ensuremath{\stylevector{c}}}
\newcommand{\ceps}{\ensuremath{\stylevector{\epsilon}}}

\newcommand{\einsmat}{\stylematrix{1}}
\newcommand{\nullmat}{\stylematrix{0}}


\newcommand{\diff}{\mathrm{d}}
\newcommand{\diffdt}[1]{\frac{\mathrm{d}#1}{\mathrm{d}t}}
\newcommand{\lie}[2]{\mathcal{L}_{#1}{#2}} 
\def\argmin{\mathop{\rm argmin} \limits} 		
\newcommand{\ddt}[0]{\frac{\mathrm{d}}{\mathrm{d}t}}
\renewcommand{\d}[0]{\mathrm{d}}
\newcommand{\dt}[0]{\mathrm{d}t}
\newcommand{\const}[0]{\mathrm{const}}
\renewcommand{\epsilon}{\varepsilon}
\newcommand{\bmat}[1]{\begin{bmatrix} #1 \end{bmatrix}}
\newcommand{\halmos}{\rule{2.0mm}{2.0mm} }
\newcommand{\laplace}{\circ\hspace{-0.15cm}-\hspace{-0.15cm}-\hspace{-0.15cm}\bullet\:}

\newcommand{\partdiff}[2]{\ensuremath{\frac{\partial#1}{\partial#2}}}
\newcommand{\RN}[1]{\textup{\uppercase\expandafter{\romannumeral#1}}}
\newcommand{\linpoint}[2]{\ensuremath{\left. #1 \right|_{#2}}}
\newcommand{\norm}[1]{\left\lVert#1\right\rVert}


\newcommand{\ol}{\overline}
\newcommand{\ds}{\displaystyle}
\def\ODelta{\mathop{\Delta}\limits}
\newcommand{\bm}[1]{\mbox{\boldmath ${#1}$}}
\newcommand{\ve}[1]{\underline{#1}}\newcommand{\ma}[1]{#1}


\newcommand{\cand}{\ensuremath{\textnormal{and}\,}}	
\newcommand{\cwith}{\ensuremath{\textnormal{with}\,}}	
\newcommand{\cmit}{\ensuremath{\textnormal{mit}\,}}	
\newcommand{\csign}{\ensuremath{\,\textnormal{sign}}}	
\newcommand{\cfor}{\ensuremath{\textnormal{for}\,}}	\newcommand{\cfuer}{\ensuremath{\textnormal{für}\,}}	\newcommand{\cund}{\ensuremath{\textnormal{und}\,}}	
\newcommand{\ckonstant}{\ensuremath{\textnormal{constant}\,}}	
\newcommand{\crange}{\ensuremath{\textnormal {range}}}
\newcommand{\crang}{\ensuremath{\textnormal {rank}}}
\newcommand{\csat}{\ensuremath{\textnormal {sat}}}
\newcommand{\cmax}{\ensuremath{\textnormal {max}}}
\newcommand{\crank}{\ensuremath{\textnormal {rank}}}
\newcommand{\cdim}{\ensuremath{\textnormal {dim}}}
\newcommand{\cdet}{\ensuremath{\textnormal {det}}}

\newcommand{\cmm}{\ensuremath{\textnormal {mm}}}

\newcommand{\RZ}{\mathbb{R}}
\newcommand{\NZ}{\mathbb{N}}
\newcommand{\CZ}{\mathbb{C}}
\newcommand{\SZ}{\mathbb{S}}
\newcommand{\SL}{\mathbb{L}}

\newcommand{\NewR}{{\rm I\hspace{-.17em}R}}
\newcommand{\NewB}{{\rm I\hspace{-.17em}B}}
\newcommand{\NewZ}{{\sf Z\hspace{-.25em}Z}}
\newcommand{\NewN}{{\rm I\hspace{-.17em}N}}
\newcommand{\Reel}{\NewR}

\newcommand{\MoAsVar}{J}


\newcommand{\fref}[1]{Abb. \ref{fig:#1}}
\newcommand{\tref}[1]{Tab. \ref{tab:#1}}
\newcommand{\mref}[1]{(\ref{eq:#1})}


\newcounter{defctr}
\renewcommand{\thedefctr}{\arabic{section}.\arabic{defctr}}
\newenvironment{definition}[1]
  	{\refstepcounter{defctr} \vspace*{1em}
    {\noindent \sffamily \bfseries{Definition \thedefctr :}} \emph{#1.} \\}

\newenvironment{motivation}
  {\centering \hspace*{0.07\textwidth}
    \begin{minipage}{0.85\textwidth} \itshape \vspace{0.4cm}}
  {\end{minipage} \vspace{1cm}}


\DeclareFixedFont{\BildFont}{OT1}{cmss}{m}{n}{10}
\newtheorem{intrem}{\hspace*{-0.8cm}$\Longrightarrow$ \footnotesize TODO}


\newcommand{\FT}{\ensuremath{F_\text{T}}}
\newcommand{\Tr}{\ensuremath{T_\text{r}}}
\newcommand{\Tg}{\ensuremath{T_\text{g}}}
\newcommand{\Jr}{\ensuremath{J_\text{r}}}
\newcommand{\Jg}{\ensuremath{J_\text{g}}}
\newcommand{\Pg}{\ensuremath{P_\text{g}}}
\newcommand{\Pel}{\ensuremath{P_\text{el.}}}
\newcommand{\vW}{\ensuremath{v}}
\newcommand{\Radius}{\ensuremath{R}}
\newcommand{\pitch}{\ensuremath{\beta}}
\newcommand{\Prot}{\ensuremath{P_\text{r}}}
\newcommand{\wR}{\ensuremath{\omega_\text{r}}}
\newcommand{\wg}{\ensuremath{\omega_\text{g}}}
\newcommand{\pitchr}{\ensuremath{\beta_\text{r}}}
\newcommand{\JoneDOF}{\ensuremath{J_\text{1DOF}}}
\newcommand{\nG}{\ensuremath{n_\text{G}}}
\newcommand{\kv}{\ensuremath{k_\vW}}
\newcommand{\kwR}{\ensuremath{k_{\omega_\text{r}}}}
\newcommand{\kpitch}{\ensuremath{k_{\pitch_\text{r}}}}
\newcommand{\kTg}{\ensuremath{k_{T_\text{g}}}}
\newcommand{\dS}{\ensuremath{d_{\text{S}}}}
\newcommand{\kS}{\ensuremath{k_{\text{S}}}}

\newcommand{\cMlong}{\ensuremath{c_\text{Q}\bigl(\lambda(\omega_\Radius,\vW),\pitch_\text{r}\bigr)}}
\newcommand{\cPlong}{\ensuremath{c_\text{P}\bigl(\lambda(\omega_\Radius,\vW),\pitch_\text{r}\bigr)}}
\newcommand{\cSlong}{\ensuremath{c_\text{T}\bigl(\lambda(\omega_\Radius,\vW),\pitch_\text{r}\bigr)}}

\newcommand{\cMshort}{\ensuremath{c_\text{Q}}}
\newcommand{\cPshort}{\ensuremath{c_\text{P}}}
\newcommand{\cSshort}{\ensuremath{c_\text{T}}}

\newcommand{\tauwr}{\ensuremath{\tau_{\omega_\text{r}}}}
\newcommand{\tauTg}{\ensuremath{\tau_{T_\text{g}}}}

\fancyhf{} 
\fancyfoot[C]{\thepage} 
\maketitle
\thispagestyle{fancy}
\pagestyle{plain}



\begin{abstract}

\noindent This paper presents a novel Model Reference Control (MRC) approach for wind turbine (WT) systems in the full load region employing a fuzzy Parallel Distribution Compensation Controller (PDC-C) derived using a Takagi-Sugeno (TS) fuzzy System approach. Through first-order Taylor series expansion, local linear submodels are generated and combined via triangular membership functions to develop a TS descriptor model. From here, the MRC PDC-C is synthesized by a constrained LMI optimization procedure, including damping characteristics of the elastic drive train, to track the desired rotor speed and generator torque based on the reference model dynamics. The controller is tested on the nonlinear WT model in simulation studies under various wind conditions, such as turbulent wind, wind gusts, and a Fault Ride Through (FRT) scenario where the generator torque is set to 0 p.u. for 150 ms.
\end{abstract}



\section{Introduction}
\noindent This work is concerned with the design of a controller for a Wind Turbine (WT) system as part of a Dynamic Virtual Power Plant (DVPP) as proposed in \cite{DVPP_Concept} and later specified through \cite{Haberle2022}. In short, the DVPP can contribute to grid ancillary services as a classical synchronous generator would, incorporating different Renewable Energy Sources (RES) as well as non-RES sources and coordinating dispatched power. In general, the studies of the DVPP approach incorporate reduced system dynamic models for the RES power plants as in \cite{Haberle2022}, \cite{Bjoerk22}, and \cite{Bjoerk23}. Therefore, the DVPP relies, to a certain degree, on the linear behavior of the underlying RES. As WT systems represent highly nonlinear systems due to the aerodynamic behavior of the rotor, a linear behavior is not ensured. To achieve the required closed-loop dynamics, a Model Reference Control (MRC) approach for the NREL 5$\,$MW WT is developed. For controller synthesis, a Takagi-Sugeno (TS) Fuzzy description model of the WT is derived and introduced in the MRC approach. A Parallel Distributed Compensation Controller (PDC-C) is developed to match the nonlinear WT behavior to the linear reference model. Furthermore, elastic drive train dynamics are incorporated into the controller design procedure, enforcing desired damping characteristics on the drive train. To verify the capability of the MRC PDC-C it is tested on the nonlinear WT model in simulation studies under various wind conditions as well as a Fault Ride Through (FRT) scenario.\\


\section{Methods}
\label{section:Methods}
\noindent \textit{Notation Remark:} In this paper matrices are denoted in bold $\cA{}$, vectors bold italic $\stylevector{x}$  and scalars italic $s$. Furthermore, $\prec$ and $\succ$ indicate negative and positive definiteness, respectively.

\subsection{Model Reference Control in TS Form}
\begin{figure}
\begin{tikzpicture}[auto, line cap=rect]

\def\xoffs{0.75}

\tikzset{packet/.style={rectangle, draw, very thick, minimum height=1.75cm,minimum width = 3.25cm}}
\tikzset{packet2/.style={rectangle, draw, very thick, minimum size=1cm}}
\tikzset{packetintegrator/.style={rectangle, draw, very thick, minimum size=0.5cm}}
\tikzset{packet3/.style={rectangle, draw, very thick, minimum width=3cm}}
\tikzset{emptynode/.style={draw=none}}
\tikzset{sum/.style={circle, draw, very thick, minimum size=0.2cm}}

    \node[packet] (Plant) at (4.5,3) {$\begin{matrix}
    \dot{\cx}=\stylevector{f}(\cx,\cu) \\ \cy=\stylevector{h}(\cx)\end{matrix}$};
    \node[packet] (Ref) at (4.5,5) {$\begin{matrix}
    \dot{\cx}^\text{r}= \cAr\cxr+\cEr\cw \\ \cyr=\cCr\cxr + \cFr\cw\end{matrix}$};
    \node[packet2] (PDCx) at (4.5,0) {$-\sum_{i=1}^{N_r}h_i(z)[\underbrace{\cK_{\cx,i},\cK_{\cxr,i},-\cK_{\mathrm{I},i}}_{\cKibar}]$};
    

    \node[packetintegrator] (integrator) at (10.5-\xoffs,2) {$\int$};
    \node[sum] (err_sum) at (10.5-\xoffs,3.25) {};
    \node[emptynode] (Refin) at (1.5,5) {};

    \node[emptynode] (muxc) at (9.625-\xoffs,0.75) {};
    \node[emptynode] (muxl) at (8.75-\xoffs,0.75) {};
    \node[emptynode] (muxr) at (10.5-\xoffs,0.75) {};

    \node[emptynode] (inp) at (1.5,3) {};
    
    \draw[line,ultra thick] (8.5-\xoffs,0.75) -- (10.75-\xoffs,0.75);
    \path [line,thick] (PDCx.west) -| (inp.center);
    \path [arrow,thick] (inp.center) -- node[above] {$\cu$} (Plant.west);
    \path [arrow,thick] (Refin.center) -- node[] {$\cw$} (Ref.west);
    \path [arrow,thick] ([yshift=0.25cm]Plant.east) -- node[at end,xshift=-0.3cm] {\cy} node[at end,below,xshift=-0.2cm] {\textbf{-}} (err_sum.west);
    
    \path [arrow,thick] ([yshift=-0.25cm]Plant.east) -| node[at end,yshift=0.4cm] {\cx} (muxl.center);
    \path [arrow,thick] (integrator.south) -- node[at end,yshift=0.4cm] {$\cx_\text{I}$} (muxr.center);
    \path [arrow,thick] ([yshift=-0.25cm]Ref.east) -| node[at end,yshift=0.45cm] {\cxr} (muxc.center);

    \path [arrow,thick] ([yshift=0.25cm]Ref.east) -| node[at end,yshift=0.3cm,xshift=0.05cm] {\cyr} (err_sum.north);
    \path [arrow,thick] (err_sum.south) -- node[] {\ceps} (integrator.north);
    
    \node[emptynode] (ref1) at (9.5,6) {};

    \path [arrow,thick] (muxc.center) |- node[at end,xshift=1cm] {$\Bar{\cx}=\begin{bmatrix}\cx\\\cxr\\\cx_\text{I}\end{bmatrix}$} (PDCx.east);

    


    \node[emptynode] (KI1) at (9.5,0) {};
    \node[emptynode] (KI2) at (2.5,0) {};

    \node[emptynode] (Kx1) at (7.5,1.5) {};

    \node[emptynode] (Kxr1) at (7.5,4.5) {};

    \node[emptynode] (Refin) at (5.5,6) {};


    \node[emptynode] (zin) at (4.5,-1.5) {};

    \path [arrow,thick] (zin.center) -- node[] {$\cz$} (PDCx.south);
    
\end{tikzpicture}
\centering
\caption{Basic PDC control structure of the MRC approach. $\cx$ and $\cxr$ denote the plant and reference model states respectively whereas $\cy$ and $\cyr$ describe the output. $\ceps=\cyr-\cy$ denotes the tracking error incorporated in the MRC state vector through integration then denoted as $\cx_\text{I}$. $\cu$ is the plant input and $\cz$ the premise variables. $\cw$ is the reference signal to the reference model. $\cKibar$ is the MRC PDC-C.}
\label{fig:MRC_cntrl_structure}
\end{figure}
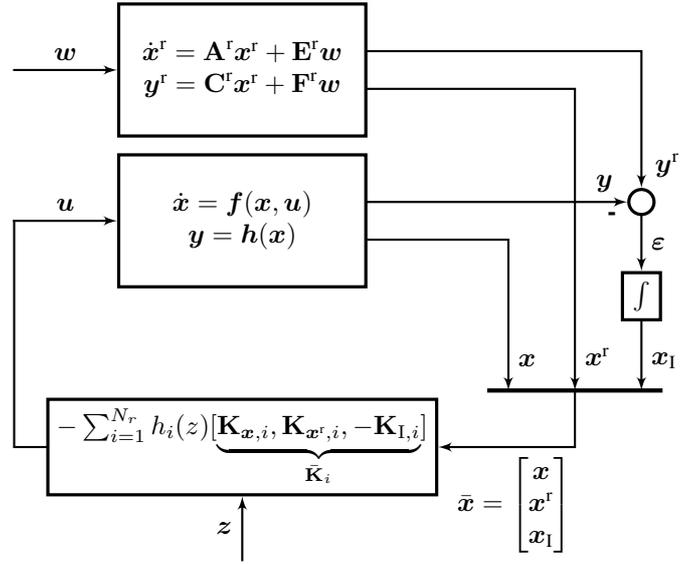
The PDC controller structure of the Model Reference Control approach is depicted in Figure \ref{fig:MRC_cntrl_structure}.
The goal of the MRC is that a given nonlinear system
\begin{equation}
    \begin{aligned}
        \dot{\cx} & = \stylevector{f}(\cx,\cu)\, ,\\
        \cy & = \stylevector{h}(\cx) \, , 
    \end{aligned}
    \label{eq:nonlin_systme}
\end{equation}
where $\stylevector{f}(\cx,\cu)$ and $\stylevector{h}(\cx)$ are vectors of nonlinear differential equations, matches the behavior of a linear reference model
\begin{equation}
\begin{aligned}
    \dot{\cx}^\text{r} & = \cAr\cxr+\cEr\cw \, ,\\
    \cyr & = \cCr\cxr + \cFr\cw \, ,
    \end{aligned}
    \label{eq:Ref_mdl_lin_ol}
\end{equation}
where $\cxr\in\RZ^{l\times1}$, $\cw\in\RZ^{p\times1}$, $\cAr\in\RZ^{l\times l}$, $\cEr\in\RZ^{l\times p}$, $\cyr\in\RZ^{q\times 1}$, $\cCr\in\RZ^{q\times l}$ and $\cFr\in\RZ^{q\times p}$. The desired reference model dynamics can be chosen arbitrarily but should consider the limitations of the dynamics of the nonlinear system. For controller synthesis (\ref{eq:nonlin_systme}) gets converted into a TS system by linearizing around selected equilibrium points $\cc_{0,i}$.
\begin{equation}
\begin{aligned}
        \cAi = &  \left. \frac{\partial\stylevector{f}}{\partial\cx} \right|_{\cc_{0,i}} \hspace{.5cm}
        \cBi = \left. \frac{\partial\stylevector{f}}{\partial\cu} \right|_{\cc_{0,i}}\hspace{.5cm}
        \cEi = \left. \frac{\partial\stylevector{f}}{\partial\cw} \right|_{\cc_{0,i}} \\
        \hspace{-.5cm}\cCi = & \left. \frac{\partial\stylevector{h}}{\partial\cx} \right|_{\cc_{0,i}}\hspace{.5cm}
        \cFi = \left. \frac{\partial\stylevector{h}}{\partial\cw} \right|_{\cc_{0,i}}
    \label{eq:JacobiAMatrix}  \, ,
\end{aligned}
\end{equation}
yielding 
\begin{equation}
    \begin{aligned}
        \dot{\cx} & = \sum\limits_{i=1}^{N_r} h_i(\cz)\big(
        \cAi{}\Delta\cx_i+\cBi{}\Delta\cu_i+\cEi{}\Delta\cw_i\big)\, ,\\
        \cy & = \sum\limits_{i=1}^{N_r} h_i(\cz)\big(\cCi{}\Delta\cx_i+\cFi{}\Delta\cw_i) \, ,
    \end{aligned} 
    \label{eq:Cont_TS_System}
\end{equation}
 where $\cx\in\RZ^{n\times1}$, $\cu\in\RZ^{m\times 1}$, $\cy\in\RZ^{q\times 1}$, $\cAi\in\RZ^{n\times n}$, $\cBi\in\RZ^{n\times m}$, $\cCi\in\RZ^{q\times n}$, $\cEi\in\RZ^{n\times p}$, $\cFi\in\RZ^{q\times p}$ with
 \begin{equation}
\begin{aligned}
   \Delta\cx_i & = \cx - \cx_{\cc_{0,i}} \, , \quad
   \Delta\cu_i = \cu - \cu_{\cc_{0,i}} \\
   \Delta\cw_i & =  \cw - \cw_{\cc_{0,i}}  \, .
\end{aligned}
\end{equation}
After a simple calculation, we obtain the equivalent affine TS model with a common state and input vector
\begin{equation}
    \begin{aligned}
        \dot{\cx} & = \sum\limits_{i=1}^{N_r} h_i(\cz)\big(
        \cAi{} \cx+\cBi{}\cu+\cEi{} \cw+ \cai\big)\, , \\
        \cy & = \sum\limits_{i=1}^{N_r} h_i(\cz)\big(\cCi{}\cx+\cFi{}\cw+\cci\big) \, , 
    \end{aligned}
    \label{eq:Cont_TS_System_with_affine_terms}
\end{equation} 
where $\cai\in\RZ^{n\times 1}$ and $\cci\in\RZ^{q\times 1}$ denote the affine terms determined by
\begin{equation}
\begin{aligned}
    \cai & = - \cAi \cx_{\cc_{0,i}} - \cBi \cu_{\cc_{0,i}} -  \cEi \cw_{\cc_{0,i}}\, , \\
    \cci & = - \cCi \cx_{\cc_{0,i}} - \cFi \cw_{\cc_{0,i}} \, .
\end{aligned}
\end{equation}
In further consideration, the affine components are neglected for the controller design. A formal proof utilizing the input-to-state stability (ISS) theorem \cite{SontagWang.1995}, \cite{SontagWang.1999} that the stability of the proposed controller is also guaranteed for affine systems is given in \cite{SchulteKusche+2023+891+908}.\\

The equations (\ref{eq:Ref_mdl_lin_ol}) and (\ref{eq:Cont_TS_System}) are incorporated into an augmented TS state-space model further denoted with an over-line bar. The augmented model incorporates the tracking error $\ceps=\dot{\cx}_\text{I}=\cyr-\cy$ between the linear reference model and the nonlinear system as well, yielding the state vector of the augmented model $\cxbar=[\cx^\top,\;\cxr{}^\top,\;\cxI^\top]^\top$. The full open-loop augmented TS model is then defined as
\begin{subequations}
    \begin{align}
    \dot{\cxbar} & = \hi 
    \begin{Bmatrix}
    \underbrace{
    \begin{bmatrix}
        \cAi & \nullmat & \nullmat \\
        \nullmat & \cAr & \nullmat \\
        -\cCi & \cCr & \nullmat 
    \end{bmatrix}}_{\cAibar}
    \cxbar+
    \underbrace{
    \begin{bmatrix}
        \cBi \\
        \nullmat\\
        \nullmat
    \end{bmatrix}}_{\cBibar}
    \cu \\ +
    \underbrace{
    \begin{bmatrix}
        \cEi \\
        \cEr \\
        \cFr - \cFi
    \end{bmatrix}}_{\cEibar}
    \cw
    \end{Bmatrix}\label{eq:MRC_TS_full_1} \\ 
    \ceps & = \hi
    \begin{Bmatrix}
    \underbrace{
    \begin{bmatrix}
        -\cCi & \cCr & \nullmat
    \end{bmatrix}}_{\cCibar}
    \cxbar + 
    \underbrace{
    \begin{bmatrix}
        \cFr-\cFi
    \end{bmatrix}}_{\cFibar}
    \cw
    \end{Bmatrix}
    \end{align} \label{eq:MRC_TS_full_2}
    \label{eq:MRC_TS_full}
\end{subequations}
with the associated full state control law in PDC form
\begin{equation}
    \cu = -\hi
    \underbrace{
    \begin{bmatrix}
        \cK_{\cx,i} & \cK_{\cxr,i} & -\cK_{\cxI,i}
    \end{bmatrix}}_{\cKibar}
    \underbrace{
    \begin{bmatrix}
        \cx \\
        \cxr \\
        \cxI
    \end{bmatrix}}_{\cxbar}
    \label{eq:TS_PDC_MRC_full_K_law}
\end{equation}
Here $\cK_{\cx,i}\in\RZ^{m\times n}$, $\cK_{\cxr,i}\in\RZ^{m\times l}$ and $\cK_{\cxI,i}\in\RZ^{m\times q}$.
For controller synthesis, the augmented model is utilized, incorporating the dynamics of the nonlinear plant represented through the local TS models, the linear reference model, and the tracking error into one single controller synthesis procedure. For the applicability of the state control law in (\ref{eq:TS_PDC_MRC_full_K_law}), the full state vector, as well as the premise variables \cz{} need to be known, either measured or observed, whereas observing the premise variable increases complexity in controller synthesis \cite{TanakaWang}. The augmented closed-loop MRC TS system is denoted as
\begin{figure}
     \centering
     \includegraphics[width=0.48\textwidth]{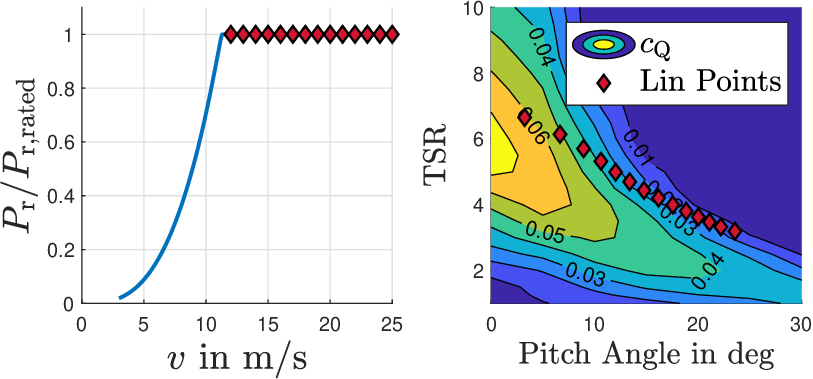}
     \caption{Linearisation points for controller synthesis on WT power curve (left) and torque coefficient surface (right).}
     \label{fig:Linpoints_Power_torque_surface}
\end{figure}
\begin{equation}
    \begin{aligned}
        \dot{\cx} & = \sum\limits_{i=1}^{N_r} \sum\limits_{j=1}^{N_r}h_i(\cz)h_j(\cz) \big(\cAibar-\cBibar\cKjbar\big) \cxbar
    \end{aligned}
    \label{eq:MRC_TS_CFS_cl_compact}
\end{equation}
The MRC approach then follows the goal to minimize the gain $\gamma$ from the change in reference signal \cw{} to the tracking error $\ceps$ defined as 

\begin{equation}
    \frac{\norm{\ceps}_2}{\norm{\cw}_2}\leq\gamma \hspace{.2cm} \iff \hspace{.2cm} \ceps^\top\ceps-\gamma^2\cw^\top\cw \leq 0 \; .
    \label{eq:minimization.}
\end{equation}

This goal is incorporated into the synthesizing of controller parameters through quadratic Lyapunov theory. To handle the possibly high amount of local models from the TS approach, an optimization procedure through LMIs, as described in \cite{TanakaWang} is favorable.

\subsection{Wind Turbine Specifications}
\begin{figure}
     \centering
     \includegraphics[width=0.45\textwidth]{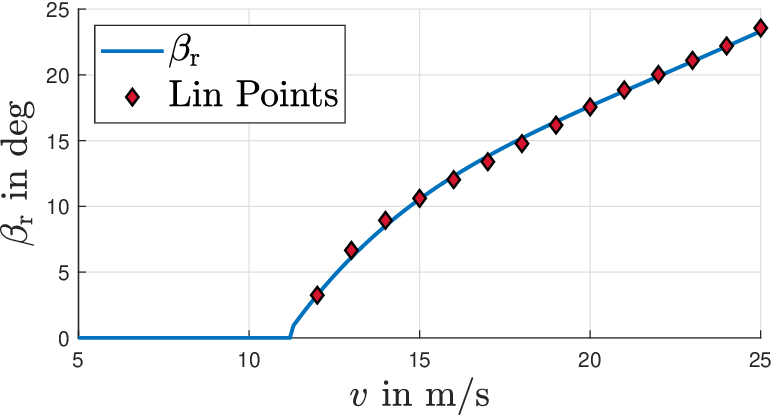}
     \caption{Linearisation points for controller synthesis on WT $\pitchr-v$ curve.}
     \label{fig:Linpoints_pitch_curve}
\end{figure}
The WT model investigated in this paper is the well-known NREL 5$\,$MW WT system given in \cite{NREL5MW}. First, the simplest representation of a WT drive train as a one-mass rotational body is derived. The nonlinear rotor torque
\begin{equation}
    \Tr=\frac{\rho}{2}\pi\Radius^3\vW^2\cMlong
    \label{eq:TorqueRotor}
\end{equation}
acts as the driving torque, and the generator torque $\Tg$ acts as the counteracting torque, resulting in the motion equation
\begin{equation}
    \dot{\wR} = \underbrace{\frac{1}{\JoneDOF}\bigl(\Tr(\vW,\wR,\pitchr)-\Tg\nG\bigr)}_{f(\vW,\wR,\pitchr,\Tg)} \, ,
    \label{eq:ODE_1massDT}
\end{equation}
with $\JoneDOF = \Jr + \nG^2 \Jg$, where $\wR$ is the rotor speed of the drive train, $\rho$ is the air density, $v$ is the wind speed far infront of the rotor, $R$ is the rotor radius, $\nG$ is the gear box ratio, $\cMshort$ is the torque coefficient, $\lambda=\frac{\wR R}{\vW}$ is the Tip Speed Ratio (TSR) and $\pitchr$ is the collective pitch angle of the rotor blades. The WT model utilizes a gearbox with $\nG$ as the gearbox ratio. Friction forces are neglected. Here, the system input of the general model \eqref{eq:Cont_TS_System_with_affine_terms}  is split into the controllable inputs of the system as $\cu = [\pitchr,\Tg]^\top$, and the disturbance defined as $d := \vW$. The state of the system is $x = \wR$. The equation (\ref{eq:ODE_1massDT}) yields the TS system based on local linearization of the form 
\begin{equation}
    \Dot{\cx} = \hi \begin{Bmatrix}\cAi \underbrace{(\wR-\omega_{\text{r},0,i})}_{x-x_{0,i}} + \cBi\underbrace{\begin{bmatrix}
    \pitchr-\beta_{\text{r,}0,i} \\
    \Tg-T_{\text{g,}0,i}
    \end{bmatrix}}_{\cu-\cu_{0,i}}
    \\
    +\cB_{\text{d}i} \underbrace{(v-v_{0,i})}_{d-d_{0,i}}\end{Bmatrix} \, .
    \label{eq:TS_1DOF_WT_DT}
\end{equation}
Operating points $\stylevector{c}_{0,i} = [x_{0,i},\cu_{0,i}^\top,d_{0,i}]^\top$ along the nominal operation regions of the WT are defined as shown in Figure \ref{fig:Linpoints_Power_torque_surface} and \ref{fig:Linpoints_pitch_curve}. This results in the submodel matrices of the TS system corresponding to the partial derivatives \eqref{eq:JacobiAMatrix} of the one-degree-of-freedom (1-DOF) model:  
\begin{equation}
    \begin{aligned}
    \cAi & = \begin{bmatrix}
    \left. \frac{\partial f(\vW,\wR,\pitchr,\Tg)}{\partial \wR{}} \right|_{\stylevector{c}_{0,i}}
    \end{bmatrix} \\
    \cBi & = \begin{bmatrix}
        \left. \frac{\partial f(\vW,\wR,\pitchr,\Tg)}{\partial\pitchr} \right|_{\stylevector{c}_{0,i}}, & \left. \frac{\partial f(\vW,\wR,\pitchr,\Tg)}{\partial\Tg} \right|_{\stylevector{c}_{0,i}}
    \end{bmatrix} \\
    \cB_{\text{d}i} & = \begin{bmatrix}\left. \frac{\partial f(\vW,\wR,\pitchr,\Tg)}{\partial\vW}\right|_{\stylevector{c}_{0,i}}\end{bmatrix}, \quad \cEi = 0 \ , \\
    \cCi & = 1 \, , \qquad \cFi = 0 \; .
    \end{aligned}  
    \label{eq:1DOF_model_matrices} 
\end{equation}
From here, the matrices in (\ref{eq:1DOF_model_matrices}) can be utilized for synthesizing controller gain sets in the MRC TS framework for rotor speed control.\\

For torque control of the generator, a stiff drive train representation is insufficient. Therefore, the physical model of the wind turbine is enhanced by an elastic two-mass model representation.
The derivation of the WT model follows the FLEX documentation \cite{FLEX}. For this purpose, the state vector of the 1-DOF model \eqref{eq:ODE_1massDT} is extended by the angular velocity of the generator and the torsion angle of the fast shaft on the generator side $\cx = [\, \wR \, ,\, \wg \, ,\, \Delta\theta \, ]^\top$. Thus, the equation of motion of the wind turbine with the drive train dynamics is given as follows 
\begin{equation}
   \label{eq:ODE_3DOF_WTmdl}
    \begin{split}   
       \dot\omega_r & = \frac{1}{\Jr} \Big( \Tr(\vW,\wR,\pitchr) - \dS \, \nG^2 \, \wR + \dS \, \nG \, \wg  -  \kS \, n_G^2 \, \Delta\theta \Big) \\
       \dot\omega_g & = \frac{1}{\Jr} \Big( \dS \, \nG \, \wR - \dS \, \wg +   \kS \, \nG \, \Delta\theta - \nG \, \Tg \Big) \\
       \Delta\dot\theta & = \wR - \frac{1}{\nG} \wg
    \end{split}   
\end{equation}
By calculating the Jacobian matrices from \eqref{eq:JacobiAMatrix} for \eqref{eq:ODE_3DOF_WTmdl}, we obtain
\begin{equation}
    \cAi = \begin{bmatrix}
        \begin{pmatrix}
            \left. \frac{1}{J_r} \frac{\partial T_r(\vW,\wR,\pitchr)}{\partial \wR{}} \right|_{\stylevector{c}_{0i}}-\frac{\dS \nG^2}{\Jr}\end{pmatrix} & \frac{\dS\nG}{\Jr} & -\frac{\kS\nG^2}{\Jr}\\
        \frac{\dS\nG}{\Jg} & -\frac{\dS}{\Jg} & \frac{\kS\nG}{\Jg}\\
        1 & -\frac{1}{\nG} & 0
    \end{bmatrix}
    \label{eq:TS_3DOF_WT_DT}
\end{equation}
for the TS submodel matrix $\cAi$, which only has a nonlinearity in the top left entry. $\dS$ and $\kS$ represent the torsional damping and spring constants respectively, whereas $\Jr$ and $\Jg$ represent the rotor and generator inertia respectively. WT model specifications are in Table \ref{tab:triebstrangkonstanten}. The input matrix is augmented to
\begin{equation}
    \cBi = \begin{bmatrix}
       \left. \frac{1}{J_r} \frac{\partial T_r(\vW,\wR,\pitchr)}{\partial\pitchr} \right|_{\stylevector{c}_{0i}} & 0  \\
       0 & - \frac{\nG}{\Jg} \\
       0 & 0
    \end{bmatrix}
    \label{eq:3DOF_Input_Matrix}
\end{equation}
and the disturbance matrix becomes
\begin{equation}
    \cB_{\text{d}i} =\begin{bmatrix}
        \left. \frac{1}{J_r} \frac{\partial \Tr(\vW,\wR,\pitchr)}{\partial\vW}\right|_{\stylevector{c}_{0i}}\\
        0\\
        0
    \end{bmatrix} \, .
\end{equation}
The remaining matrices from \eqref{eq:JacobiAMatrix} are 
\begin{equation}
    \cEi  = \begin{bmatrix} 
                0&0\\
                0&0\\
                0&0\\
            \end{bmatrix}  , \quad 
    \cCi = \begin{bmatrix}
              1&0&0\\
              0&0&0
         \end{bmatrix}    , \quad 
    \cFi = \begin{bmatrix} 
                0&0\\
                0&0\\
            \end{bmatrix} \; .
\end{equation}
The extended model allows the incorporation of the elastic properties of the drive train into the controller synthesis procedure. Desired characteristics such as damping on the drive train states can, therefore, be enforced by simultaneously respecting the limiting nonlinear dynamics of the WT system through constraints on the controller synthesis procedure.\\

To apply the MRC approach to the torque control of the WT System, a further augmentation to (\ref{eq:TS_3DOF_WT_DT}) must be conducted. By incorporating an actuator dynamic for the generator torque with a time constant $\tau_{T_\text{g,act}}$ and a demanded torque $T_\text{g,d}$ as
\begin{equation}
    \dot{T}_\text{g} = -\frac{1}{\tau_{T_\text{g,act}}}\Tg + \frac{1}{\tau_{T_\text{g,act}}}\Tg{}_\text{,d} \;. 
    \label{eq:GenTorque_Dynamic}
\end{equation}
The state transition matrix (\ref{eq:TS_3DOF_WT_DT}) becomes
\begin{equation}
\fontsize{9pt}{12pt}\selectfont
    \cAi = \begin{bmatrix}
        \begin{pmatrix}
            \left.  \frac{1}{J_r} \frac{\partial T_r(\vW,\wR,\pitchr)}{\partial \wR{}} \right|_{\stylevector{c}_{0,i}}-\frac{\dS \nG^2}{\Jr}\end{pmatrix} & \frac{\dS\nG}{\Jr} & -\frac{\kS\nG^2}{\Jr} & 0\\
        \frac{\dS\nG}{\Jg} & -\frac{\dS}{\Jg} & \frac{\kS\nG}{\Jg} & -\frac{\nG}{\Jg}\\
        1 & -\frac{1}{\nG} & 0 & 0\\
        0 & 0 & 0 & -\frac{1}{\tau_{T_\text{g,act}}}
    \end{bmatrix}
    \label{eq:TS_3DOF_TgDynamic}
\end{equation}
with the input matrix
\begin{equation}
    \cBi = \begin{bmatrix}
        \left. \frac{1}{J_r} \frac{\partial T_r(\vW,\wR,\pitchr)}{\partial\pitchr} \right|_{\stylevector{c}_{0,i}}&0  \\
        0&0 \\
        0&0\\
        0&\frac{1}{\tau_{T_\text{g,act}}}
    \end{bmatrix} \, .
    \label{eq:TS_3DOF_TgDynamic_Input}
\end{equation}
The state vector of the system now includes the current generator torque $\cx = [\wR,\wg,\Delta\theta,\Tg]^\top$ and can be defined as an output of the system $\cy = [\wR,\Tg]^\top$ through the output matrix

\begin{equation}
    \cCi = \begin{bmatrix}1&0&0&0\\
    0&0&0&1\end{bmatrix}
    \label{eq:TS_3DOF_TgDynamic_Output}
\end{equation}\\
and later be compared to a reference value $\cyr = [\wR{}_{\text{,ref}},\Tg{}_{\text{,ref}}]^\top$. The input vector changes to $\cu=[\pitchr,\Tg{}_{\text{,d}}]^\top$. This approach allows $\Tg$ and $\wR$ to be simultaneously referenced to a desired behavior through the reference model. Note, as before with the third-order system, the matrices $\cEi$, $\cFi$ according to \eqref{eq:JacobiAMatrix} are zero matrices. 

\subsection{Constraint LMI Optimization for TS MRC Approach}
\label{section:optimization_procedure}
The optimization procedure is based on the quadratic Lyapunov approach, with the function candidate $V(\cx)$ and its derivative $\dot{V}(\cx)$

\begin{subequations}
\label{eq:lyapunovfunction}
    \begin{align}
    V(\cx) & = \cx^\top \stylematrix{P}\cx
    \label{eq:Quadratic_Lyapunov_Function}\\
    \dot{V}(\cx) & = \dot{\cx}^\top \stylematrix{P}\cx+\cx^\top \stylematrix{P}\dot{\cx}\label{eq:diff_quadratic_lyapunov}
    \end{align}
\end{subequations}\\
which must fulfill the properties 

\begin{subequations}
\label{eq:lyapunov_properties-all}
    \begin{align}
         V(\cx) &> 0,\;\;\forall\cx\neq0\label{eq:lyapunov_properties-1}\\
         V(\cx) &= 0,\;\;\cx=0\label{eq:lyapunov_properties-2}\\
         \dot{V}(\cx) &< 0,\;\;\forall\cx\neq0\label{eq:lyapunov_properties-3}
    \end{align}
\end{subequations}\\
With (\ref{eq:MRC_TS_CFS_cl_compact}), (\ref{eq:minimization.}),(\ref{eq:lyapunovfunction}) and (\ref{eq:lyapunov_properties-all}) an LMI optimization formulation can be derived including constraints on the system dynamics through a so-called $\mathcal{D}$-Region, depicted in Figure \ref{fig:D_region_alpha_min_max_tikz}. The goal of the optimization problem is formulated to minimize the gain $\gamma$ from reference signal $\cw$ to tracking error $\ceps$ under the constraints:
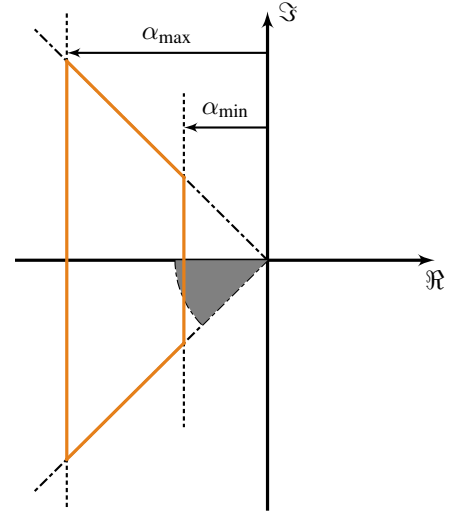
\begin{figure}
\centering
\begin{tikzpicture}[scale=2.2,line cap=rect]
    \tikzstyle{arrow} = [draw, -latex']
    \tikzset{emptynode/.style={draw=none}}
    \def\decayrate{0.5} 
    \def\decayratemax{1.2}
    \def\radius{1.4}    
    \def\angleCone{45}  
    \def\col{orange!80!gray}

    \draw[arrow,very thick] (-1.5,0) -- (1,0) node[below] {$\Re$};
    \draw[arrow,very thick] (0,-1.5) -- (0,1.5) node[right] {$\Im$};

    \draw[thick,dashdotted] (0,0) -- (-\radius, {\radius*tan(\angleCone)});
    \draw[thick,dashdotted] (0,0) -- (-\radius, {-\radius*tan(\angleCone)});
    \node[emptynode] (angle) at (-0.28,-0.12) {$\theta$};
    \draw[thick,dashdotted] (180:\radius-0.85) arc (180:180+\angleCone:\radius-0.85);
    \fill[gray, opacity=0.3] (180:\radius-0.85) arc (180:180+\angleCone:\radius-0.85) -- (0,0) -- cycle;

    \draw[very thick,\col] (-\decayrate,{\decayrate*tan(\angleCone)}) -- (-\decayrate,{-\decayrate*tan(\angleCone)});
    \draw[thick,dotted] (-\decayrate,{\decayrate*tan(\angleCone)}) -- (-\decayrate,1);
    \draw[thick,dotted] (-\decayrate,{-\decayrate*tan(\angleCone)}) -- (-\decayrate,-1);
    \draw[thick,arrow] (0,0.8) -- node[above] {$\alpha_\text{min}$} (-\decayrate,0.8);

    \draw[very thick,\col] (-\decayratemax,{\decayratemax*tan(\angleCone)}) -- (-\decayratemax,{-\decayratemax*tan(\angleCone)});
    \draw[thick,dotted] (-\decayratemax,{\decayratemax*tan(\angleCone)}) -- (-\decayratemax,1.5);
    \draw[thick,dotted] (-\decayratemax,{-\decayratemax*tan(\angleCone)}) -- (-\decayratemax,-1.5);
    \draw[thick,arrow] (0,1.25) -- node[above] {$\alpha_\text{max}$} (-\decayratemax,1.25);

    \draw[very thick,\col] (-\decayrate,{\decayrate*tan(\angleCone)}) -- (-\decayratemax,{\decayratemax*tan(\angleCone)});
    \draw[very thick,\col] (-\decayrate,{-\decayrate*tan(\angleCone)}) -- (-\decayratemax,{-\decayratemax*tan(\angleCone)});
    
\end{tikzpicture}
\caption[$\mathcal{D}$-Region representation in the complex plane]{$\mathcal{D}$-Region representation in the complex plane for a minimum and maximum decay constraint $\alpha_\text{min}$ and $\alpha_\text{max}$ and a desired damping through a cone with angle $\theta$.}
\label{fig:D_region_alpha_min_max_tikz}
\end{figure}

\begin{figure*}[t]
     \centering
    \includegraphics[width=1\textwidth]{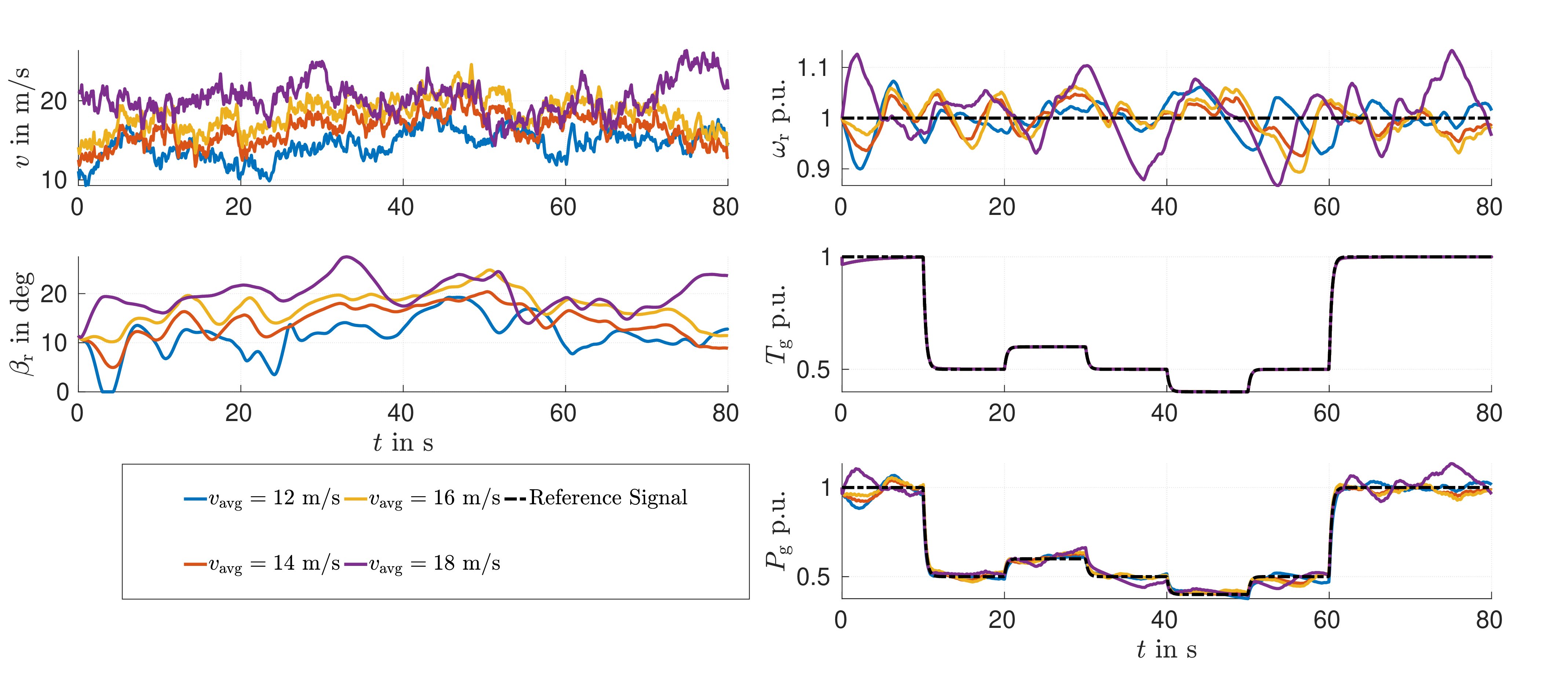}
     \caption{WT behavior under MRC PDC-C in different turbulent wind conditions with change in reference torque under constant rotational speed reference. The colors of the plot lines relate to the turbulent wind speed.}
     \label{fig:turb_testcase_3}
\end{figure*}

\begin{table}[]
\centering
\caption{Controller Synthesis LMI Constraints and Reference Model Parameters}
\begin{tabular}{ccc}
\toprule
\multicolumn{3}{c}{\textbf{Constraints}}\\[5pt]
 & $\cKjbar{}_{,\wR{}}$   & $\cKjbar{}_{,\Tg{}}$\\
\midrule
$\alpha_\text{min}$ & 0.10  & 0.20 \\
$\alpha_\text{max}$ & 1.00  & 2.00 \\
$\theta$ in rad    & 1.51  & 1.51 \\
$\gamma$      & 3.00  & 1.50 \\
$\tau_{\Tg{}}$     &  /     & 0.30 \\
$\tau_{\wR{}}$     & 10.00 & 4.00\\
\bottomrule
\end{tabular}
\label{tab:controller_constraints}
\end{table}

\vspace{-.5cm}
\begin{subequations}
\begin{align}
      &\hspace{-7cm}\min_{\cX,\;\cMjbar}\gamma\nonumber\\
      &\hspace{-7cm}\text{subject to}  \nonumber \\
    \cX\succ0,\\[5pt]
    \cX\cAibar^\top+\cAibar\cX-\cMjbar^\top\cBibar^\top-\cBibar\cMjbar+2\alpha_\text{min}\cX\prec0, \\[5pt]
    \cX\cAibar^\top+\cAibar\cX-\cMjbar^\top\cBibar^\top-\cBibar\cMjbar+2\alpha_\text{max}\cX\succ0,\\[5pt]
    \begin{aligned} &
        \left[\begin{matrix}
          \sin{\theta}(\cX\cAibar^\top+\cAibar\cX-\cMjbar^\top\cBibar^\top-\cBibar\cMjbar)\\
          \cos{\theta}(\cX\cAibar^\top - \cMjbar^\top\cBibar^\top - \cAibar\cX +\cBibar\cMjbar)
        \end{matrix}\right.\\
        &\qquad
        \left.\begin{matrix}
          \cos{\theta}(\cAibar\cX- \cBibar\cMjbar - \cX\cAibar^\top + \cMjbar^\top\cBibar^\top )\\
          \sin{\theta}(\cAibar\cX - \cBibar\cMjbar + \cX\cAibar^\top - \cMjbar^\top\cBibar^\top )
        \end{matrix}\right]
    \end{aligned}\prec 0 \\[5pt]
    \begin{bmatrix}
        \begin{pmatrix}
            \hspace{-0.6cm}\cX\cAibar^\top+\cAibar\cX \\
            \hspace{0.1cm}-\cBibar\cMjbar-\cMjbar^\top\cBibar^\top
        \end{pmatrix} & \cEibar & \cX\cCibar^\top \\
        \cEibar^\top & -\gamma^2\cI & \cFibar^\top \\
        \cCibar\cX  & \cFibar & -\cI
    \end{bmatrix}\prec0
    \end{align}
    \label{eq:MRC-D-LMI-all}
\end{subequations}
$\text{for all}\;i,j=1,\dots, N_r \; \text{s.t.} \; h_i \cdot h_j \; \neq \; 0$\\

The notation "$\text{for all}\;i,j=1,\dots, N_r \; \text{s.t.} \; h_i \cdot h_j \; \neq \; 0$" states that only TS submodels are taken into account in the LMI formulation, which can be active at the same time \cite{Poeschke2022}. It is also possible to not follow a minimization procedure and enforce a chosen $\gamma$ on the system. If $\gamma$ is chosen appropriately, a high decay on the tracking error is achieved. Due to the dynamics of the reference model not being influenced by the controllable inputs of the system, a violation of the $\mathcal{D}$-Region by the reference model dynamics may arise. The procedure where a fixed $\gamma$ is enforced on the system is conducted in this work. To solve LMIs, the open-source solver SeDuMi \cite{SeDuMi} is used. If the solver is capable of satisfying the constraints and finding a matrix $\cX$ and $N_r$ matrices $\cMjbar$, the gain parameters for each linearization point can be obtained through $\cKjbar = \cMjbar\cX^{-1}$.

\subsection{Controller Design Specifications}
To decouple the constraints through the $\mathcal{D}$-Region on the rotor speed $\wR$ and generator torque $\Tg$ two controller synthesis procedures are followed and later combined for the full MRC PDC-C. The constraint parameters are listed in Table \ref{tab:controller_constraints}. It is noted here that the matrices $\cEi,\; \cFi,\; \text{and}\; \cFr$ are zero for controller synthesis. Further, it is assumed that the $N_r$ pairs of TS submodels are controllable as stated in \cite{PoeschkeGauterinSchulte2019}.\\

\subsubsection{Controller for Rotational Dynamics}
For control of the rotational dynamics of the WT system, the 1$\,$DOF drive train model derived in (\ref{eq:ODE_1massDT}), (\ref{eq:TS_1DOF_WT_DT}) and (\ref{eq:1DOF_model_matrices}) is sufficient and is simplified by only actuating the collective pitch angle $\pitchr$ of the blades giving
\begin{equation}
    \cBi = \begin{bmatrix}
        \left. \frac{\partial f(\vW,\wR,\pitchr,\Tg)}{\partial\pitchr} \right|_{\stylevector{c}_{0,i}}
    \end{bmatrix} 
\end{equation}
as input matrix. As a reference model, a first-order system of the form
\begin{equation}
    \cAr = -\frac{1}{\tau_{\omega_\text{r}}},\hspace{1cm} \cEr = \frac{1}{\tau_{\omega_\text{r}}},\hspace{1cm} \cCr = 1
\end{equation}
where $\tau_{\omega_\text{r}}$ denotes the selectable time constant, is used. From here, the controller gain parameters can be synthesized through the procedure in Section \ref{section:optimization_procedure}.
The gain parameters for the rotational dynamics are further denoted as $\cKjbar{}_{,\omega_\text{r}}=\begin{bmatrix} k_{x,\wR{},j} & k_{x_\text{r},\wR{},j} & k_{\text{I},\wR{},j} \end{bmatrix}$. The eigenvalues of the closed-loop TS submodels for $\wR$ tracking are further denoted as $\lambda_{i,\wR{}}$.\\
\subsubsection{Controller for Torque Control}
The controller for torque control should incorporate elastic drive train dynamics to actively dampen the drive train states, as well as respect the dynamical limitations of the WT system and simultaneously track the desired torque from the reference signal. To achieve that, the WT system description in (\ref{eq:TS_3DOF_TgDynamic}), (\ref{eq:TS_3DOF_TgDynamic_Input}) and (\ref{eq:TS_3DOF_TgDynamic_Output}) is used. in combination with the reference model 
\begin{equation}
    \begin{matrix}
    \cAr = \begin{bmatrix}
        -\frac{1}{\tau_{\omega_\text{r}}}&0\\
        0&-\frac{1}{\tau_{T_\text{g}}}
    \end{bmatrix} &
    \cEr = \begin{bmatrix}
        \frac{1}{\tau_{\omega_\text{r}}}&0\\
        0&\frac{1}{\tau_{T_\text{g}}}
    \end{bmatrix} &
    \cCr = \begin{bmatrix}
        1&0\\
        0&1
    \end{bmatrix} 
    \end{matrix}
\end{equation}
with the time constants $\stylevector{\tau}_{\text{ref}}=[\tau_{\omega_\text{r}},\;\tau_{T_\text{g}}]$. For controller synthesis, $\tau_{T_\text{g}}=\tau_{T_\text{g,act}}$ should be considered as that allows for the exact tracking of $\Tg{}_\text{,ref}$ through $\Tg$. As for the controller for rotational dynamics, the procedure in section \ref{section:optimization_procedure} is followed to obtain the gain sets $\cKjbar{}_{,\text{T}_g}\in\RZ^{2\times 8}$. The eigenvalues of the closed-loop TS submodels for torque control are further denoted as $\lambda_{i,\Tg{}}$.\\
\subsubsection{Combination of Controller Parameters}
Due to potentially different desirable constraints on the rotational dynamics and torque dynamics of the WT system, the gain sets $\cKjbar{}_{,\omega_\text{r}}$ and $\cKjbar{}_{,\text{T}_g}$ are combined in the final gain matrix for the multi-tracking objective of the MRC PDC-C. Here the first row of the controller gains  $\cKjbar{}_{,\text{T}_g}$ is replaced with $\cKjbar{}_{,\omega_\text{r}}$ whereas the decoupling of the reference channels is achieved by adding zeros at the according positions yielding the final gain matrices
\begin{equation}
\fontsize{9pt}{12pt}\selectfont
    \cKjbar{}_{,\text{MT}}=
    \begin{bmatrix}
        \begin{bmatrix}k_{x,\wR{},j} & \nullmat^{1\times3}\end{bmatrix} & k_{\stylevector{x}^\text{r},\wR{},j} & 0 & k_{\text{I},\wR{},\text{j}} & 0 \\
        \cK^{1\times4}_{\stylevector{x},\Tg{},j} & 0 & k_{\stylevector{x}^\text{r},\Tg{},j} & 0 & k_{\text{I},\Tg{},j} \\
    \end{bmatrix}
    \label{eq:Full_MRC_K}
\end{equation}
The eigenvalues of the according closed-loop TS submodels are denoted as $\lambda_{i,\text{MT}}$. As the controller gain matrices are altered through the procedure described above, proof of stability can be obtained by proving that $\lambda_{i,\text{MT}}$ still fulfill LMI constraints as presented in \cite{Poeschke2022}.
\section{Results}
\label{section:Results}
\begin{figure}
     \centering
     \includegraphics[width=0.5\textwidth]{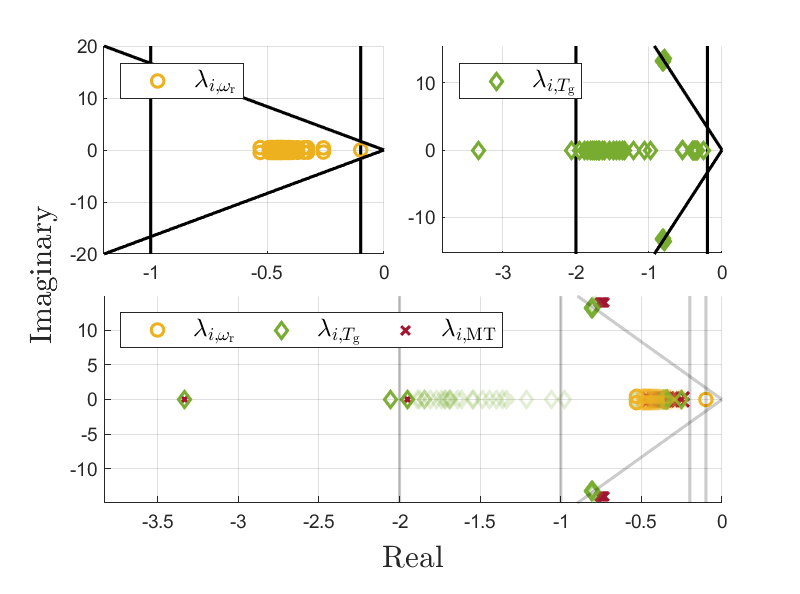}
     \caption{Eigenvalues with constraints (black lines) of closed-loop eigenvalues of the TS submodel combinations for controller synthesis of rotational dynamics $\lambda_{i,\wR{}}$, torque control $\lambda_{i,\Tg{}}$ and the resulting combined controller $\lambda_{i,\text{MT}}$}
     \label{fig:evs_MRC_MIMO_paper}
\end{figure}

The controller in (\ref{eq:Full_MRC_K}) obtained through the mixed controller synthesis approach is tested on the nonlinear model of the NREL 5$\,$MW WT system under various wind conditions. Furthermore, a short depiction of the closed-loop eigenvalues for all TS submodels, which can be active simultaneously, is shown in Figure \ref{fig:evs_MRC_MIMO_paper}. Interestingly, the eigenvalues of the tracking channels do not vary over the TS submodel combinations in $\lambda_{i,\text{MT}}$. It is noted here that to emulate an observer behavior, a first-order system from $v$ to $z$ with a time constant of 2$\,$s as well as a pitch rate limit as given in the turbine specifications in Table \ref{tab:triebstrangkonstanten} is incorporated in the simulation.
\subsection{Turbulent Wind Conditions}
Figure \ref{fig:turb_testcase_3} shows the nonlinear WT model under MRC PDC-C control for various turbulent wind fields with an average of $\stylevector{v}_\text{turb,avg}\in\{12,14,16,18\}$ m/s. At $t=10\,$s, the generator torque reference is changed to 0.5 p.u. and then varied around the new reference with $\pm$0.1 p.u. steps every 10 seconds. At $t=60\,$s the torque reference is set back to 1.0 p.u.. It is observable that the generator torque closely follows the reference. Furthermore, the change in generator power output only varies due to the change in rotational speed of the turbine resulting from the turbulent wind conditions. A deterministic pitch behavior can not be observed, as changes in the wind speed might cancel out the reduction or increase of the generator torque.
\subsection{Wind Gusts}
\begin{figure}
     \centering
     \includegraphics[width=0.5\textwidth]{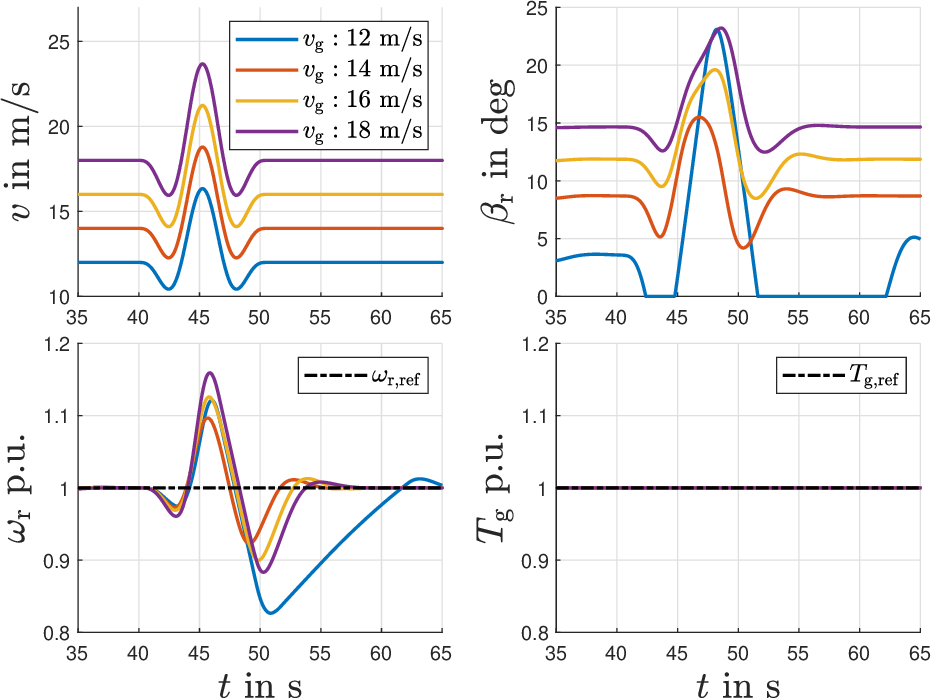}
     \caption{WT system under MRC PDC-C in various wind gust scenarios $\stylevector{v}_\text{g}\in\{12,14,16,18\}$ m/s (top left) in the full load region showing the pitch $\pitchr$ (top right) and rotor speed $\wR$ (bottom left) response under constant generator torque (bottom right) input.}
     \label{fig:gusts_paper}
\end{figure}

To see the performance of the controller robustness against strong disturbances on the WT rotor speed, a simulation with wind gusts is conducted following the DIN EN IEC 61400-1 standard \cite{DIN_EN_IEC_61400-1}. In total, four wind gusts with an initial wind speed of $\stylevector{v}_\text{g}\in\{12,14,16,18\}$ m/s are tested. The simulation results are shown in Figure \ref{fig:gusts_paper}. The generator torque is kept constant. As the lowest gust with 12 m/s initial wind speed is at the lower bound of the full load region the pitch reaches zero degree in the dip of the wind speed before the main peak of the gust therefore reaching the transition point from full to partial load region of the WT leaving the controller design range.
\subsection{Fault Ride Through Scenario}
\begin{figure*}
     \centering
    \includegraphics[width=1\textwidth]{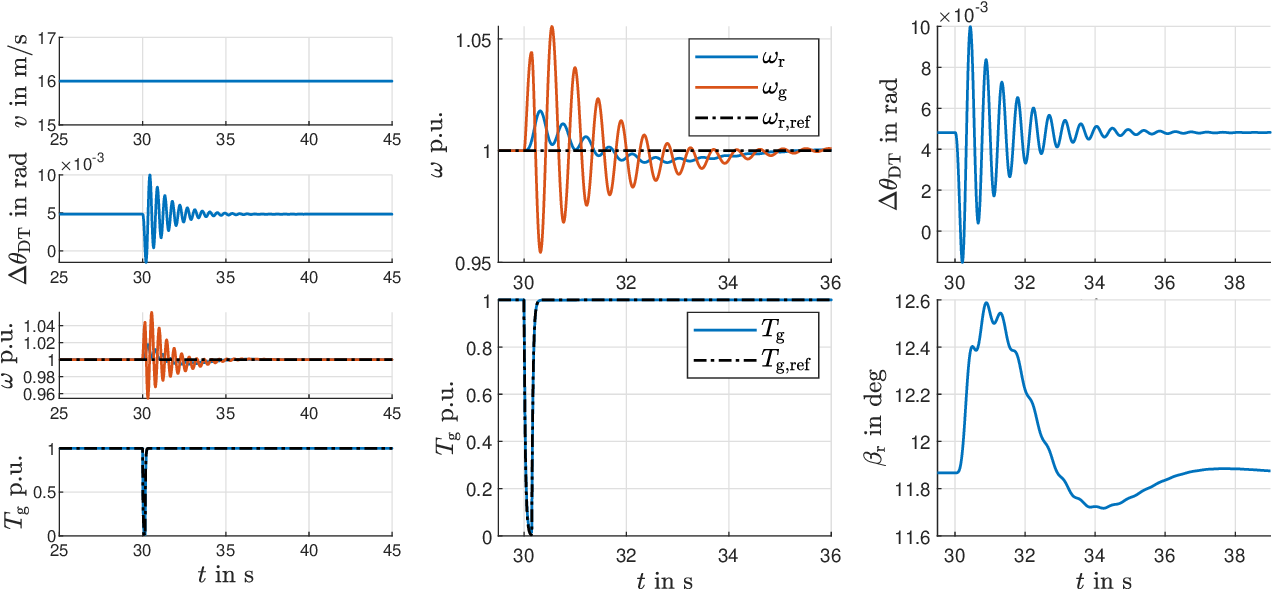}
     \caption{WT system under MRC PDC-C in FRT scenario with a $\Tg$ jump to 0 p.u. for 150 ms under constant wind speed with 16 m/s. Showing the rotor and generator speed (top center), torque input (bottom center), drive train torsional angle (top right), and pitch angle (bottom right).}     
     \label{fig:FRT_150ms}
\end{figure*}

A FRT scenario describes a voltage drop of a certain period in the grid the WT system is connected to as, i.e., defined by the European Union in \cite{NCFG}. Usually, national guidelines set requirements for how the WT system should behave during FRT, as shown in \cite{FRT2018}. Due to the coupling of the mechanical and electrical dynamics through the WT generator, a FRT scenario does have impacts on the torque and pitch control of the WT. Therefore, the FRT scenario for Germany described in \cite{FRT2018} is set as a benchmark for a test of the MRC PDC-C. The scenario describes a grid side voltage drop to 0 p.u. for 150 ms on the transmission voltage level. The turbine cannot feed active power into the grid if no grid voltage is present. As a result, the torque of the generator must be set to 0 p.u. for the time of the fault. To achieve the desired characteristics through the reference model $\tau_{\Tg{}}$ is set to 25 ms. Therefore, the generator torque reaches the desired reference torque of 0 p.u. within the 150 ms. The controller parameters are not altered from the original controller synthesis due to the linear behavior of the torque system states. A depiction of the simulation results, including the elastic drive train response, is shown in Figure \ref{fig:FRT_150ms}.
\subsection{Simultaneous Reference Change}
\begin{figure}
     \centering
     \includegraphics[width=0.41\textwidth]{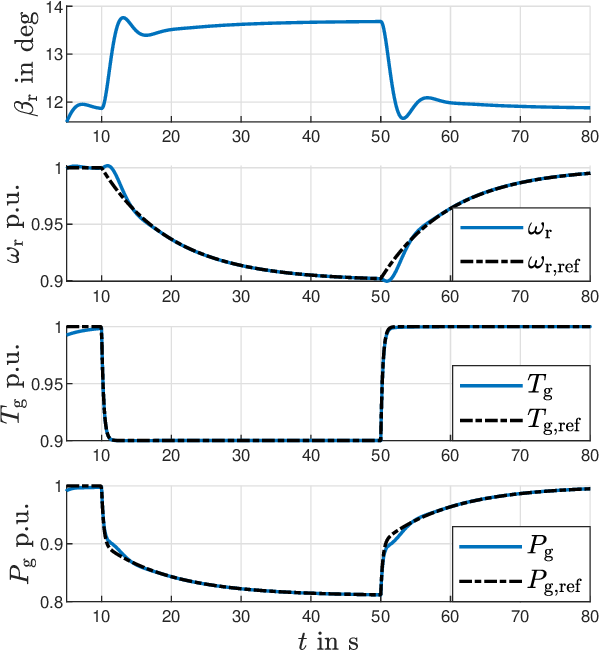}
     \caption{WT system under MRC PDC-C in constant wind speed conditions of 16 m/s with a simultaneous change in torque and rotor speed reference (middle two). Showing pitch response (top) and resulting generator power (bottom).}
     \label{fig:simult_ref_change}
\end{figure}

So far, only the change in reference value for $\Tg$ was presented. In contrast, the pitch controller was only occupied with returning the rotor speed to its nominal value as the turbulent wind speed or changes in generator torque disturbed $\wR$. The capabilities of the MRC PDC-C include changes in $\wR$ and $\Tg$ simultaneously, as shown in Figure \ref{fig:simult_ref_change}. Here, a change in reference of $\Tg$ and $\wR$ of $-0.1$ p.u. respectively occurs at $t=10\,$s and is referenced back to nominal at $t=50\,$s. As the time constants $\stylevector{\tau}_{\text{ref}}$ are chosen differently, the contribution from the change of $\Tg$ to the total power change of $-0.2$ p.u. is faster than the contribution from change in $\wR$.
\section{Discussion}
\label{section:Discussion}
The MRC PDC-C developed within this work showed promising behavior within the conducted simulation studies. Due to the TS model formulation, the approach can decouple constraints to meet different desired design goals of the WT system as similarly presented in \cite{Poeschke22}. If the nonlinear behavior of the WT is modeled sufficiently enough through the TS subsystems, analytical stability through a quadratic Lyapunov function can be proven.\\[-9pt]

The simulation studies were conducted under ideal conditions, and all states for the controller, as well as the premise variable, were measurable without noise. Usually, an observer for the unmeasurable states is developed, which can also be done in the TS framework as shown in \cite{ichalal2010} and specified for WT systems in \cite{Poeschke22}. The MRC PDC-C incorporates the wind speed as the premise variable $z=v$, which needs to be estimated. 
Various techniques exist, as shown in \cite{gauterin2016}. A reconstruction of the premise variable demands that the controller synthesis is extended to include further TS submodels into the controller synthesis, as shown in \cite{Poeschke22}. Therefore, the MRC PDC-C approach presented here should be tested under the incorporation of a nonlinear observer and sensor noise. Further, an extended controller synthesis incorporating uncertainties by a reconstructed premise variable should be conducted to assure stable behavior in realistic WT systems.\\[-9pt]

Various simulation studies of the WT system under MRC PDC-C found that increasing the rotor speed of the WT under constant torque input was limited. Investigations showed that a strong increase of $\wR$ from 0.5 p.u. to 1 p.u. with a time constant of $\tau_{\wR{}}=10\,$s and a constant wind speed of 16 m/s demanded low pitch angles $\pitchr$ with a simultaneously low TSR. Therefore, by lowering $\pitchr$, the torque coefficient first increases and then decreases the rotor torque. This can be imagined by a movement from the right-hand flank of the $c_\text{Q}$ surface in Figure \ref{fig:Linpoints_Power_torque_surface} to the left Flank, moving over the ridge of the surface, where the torque coefficient is its highest for that TSR.\\[-9pt]

The FRT simulation studies do not incorporate an electrical model in the simulation. Solely, the torque was changed through the reference channel, incorporating a lower time constant $\tau_{\Tg{}}$ for the reference model; therefore, no potential interactions between the electrical and mechanical models can be proven.\\[-9pt]

Future work should includes the implementation of sufficient observers into the simulation and the expansion of the control approach to the partial load region of the WT system. As rotor speed increases through the reference signal shown unstable behavior, the linearization points and the premise variables could be extended to incorporate sufficient knowledge of the $\cMshort$ surface. The controller, in theory, should then be capable of controlling the WT in a larger operational region. To see if the controller can perform sufficiently inside a WT system, additional studies should be conducted incorporating electrical models into the simulation. Additionally, tower and blade models should be incorporated into the simulation to see the effects of the control action on the structural elements of the WT.\\[-15pt]
\section{Conclusion}
\label{section:Conclusion}
 A novel nonlinear reference control structure through an MRC PDC-C was presented. The controller is able to control a WT system under various wind conditions in the full load operational region, naming turbulent wind, wind gusts, and an emulated FRT scenario. Furthermore, the rotor speed and generator torque could be referenced simultaneously. Through the enforced linear behavior of the nonlinear WT, the control approach can meet the DVPP conditions, which assume the near linear behavior of RES systems. Due to the flexible controller design approach, it can be extended to various RES, allowing for compact controller synthesis for RES, which is part of DVPPs.

\begin{table}
\centering
\caption{NREL 5MW Turbine Specifications \cite{NREL5MW}}
\fontsize{9pt}{15pt}\selectfont
\begin{tabular}{lcrl} 
\toprule
\textbf{Description}     & \textbf{Symbol}            & \multicolumn{1}{c}{\textbf{Value}} & \multicolumn{1}{c}{\textbf{Unit}}  \\ 
\midrule
Electrical power rating & $\Pel{}_\text{,rated}$ & 5 & MW\\
\makecell[l]{Electrical Generator \\[-5pt] Efficiency} & $\eta_\text{el.}$ & 0.944 & \\
Rated generator torque & $\Tg{}_\text{,rated}$ & 43093.55 & Nm\\
Rated generator speed & $\omega_\text{g,rated}$ & 1173.7 & rpm  \\
Rated mechanical power & $P_\text{mech,rated}$ & 5.297 & MW\\
Cut in windspeed & $v_{\text{cut in}}$ & 3 & m/s\\
Rated windspeed & $v_{\text{rated}}$ & 11.4 & m/s\\
Cut out windspeed & $v_{\text{cut out}}$ & 25 & m/s\\
Cut in rotor speed & $\wR{}_\text{,cut in}$ & 6.9 & rpm\\
Rated Rotor speed & $\wR{}_\text{,rated}$  & 12.1 & rpm \\
Gear ratio & \nG & 97 &  \\
Rotor radius & $R$ & 63 & m \\
Inertia Rotor & \Jr & 38759227 & kg m\textsuperscript{2}  \\
Inertia Generator & \Jg & 534.1  & kg m\textsuperscript{2} \\
Torsional stiffnes shaft & \kS & 92214 & Nm/rad \\
Torsional damping &   \dS & 660.54 & Nms/rad \\
Max generator Torque &  $\Tg{}_\text{,max}$ & 47402.91 & Nm \\
Max generator Torque rate &  $\Tg{}_\text{,rate,max}$ & 15000 & Nm/s \\
Max Blade-Pitch rate &  $\pitchr{}_\text{,rate,max}$ & 8 & $^\circ$/s \\
Peak Power Coefficient &  $\cPshort{}_\text{,opt}$ & 0.482 & \\
\makecell[l]{Tip-Speed-Ratio\\[-5pt] at Peak Power\\[-5pt] Coefficient} &  $\lambda_\text{opt}$ & 7.55 & \\
\bottomrule
\end{tabular}
\label{tab:triebstrangkonstanten}
\end{table}

\section*{Acknowledgment}

This research is part of the EU-Project POSYTYF
(POwering SYstem flexibiliTY in the Future through RES), https://posytyf-h2020.eu. The POSYTYF project has received funding from the European Union’s Horizon 2020  research and innovation programme under grant agreement No 883985. 

\bibliography{MA_Quellen,Lit_HS_2024_01_10}

\begin{thebibliography}{10}
\providecommand{\url}[1]{#1}
\csname url@samestyle\endcsname
\providecommand{\newblock}{\relax}
\providecommand{\bibinfo}[2]{#2}
\providecommand{\BIBentrySTDinterwordspacing}{\spaceskip=0pt\relax}
\providecommand{\BIBentryALTinterwordstretchfactor}{4}
\providecommand{\BIBentryALTinterwordspacing}{\spaceskip=\fontdimen2\font plus
\BIBentryALTinterwordstretchfactor\fontdimen3\font minus \fontdimen4\font\relax}
\providecommand{\BIBforeignlanguage}[2]{{%
\expandafter\ifx\csname l@#1\endcsname\relax
\typeout{** WARNING: IEEEtran.bst: No hyphenation pattern has been}%
\typeout{** loaded for the language `#1'. Using the pattern for}%
\typeout{** the default language instead.}%
\else
\language=\csname l@#1\endcsname
\fi
#2}}
\providecommand{\BIBdecl}{\relax}
\BIBdecl

\bibitem{DVPP_Concept}
B.~Marinescu, O.~Gomis-Bellmunt, F.~Dörfler, H.~Schulte, and L.~Sigrist, ``Dynamic virtual power plant: A new concept for grid integration of renewable energy sources,'' \emph{IEEE Access}, vol.~10, pp. 104\,980--104\,995, 2022.

\bibitem{Haberle2022}
V.~Häberle, M.~W. Fisher, E.~Prieto-Araujo, and F.~Dörfler, ``Control design of dynamic virtual power plants: An adaptive divide-and-conquer approach,'' \emph{IEEE Transactions on Power Systems}, vol.~37, no.~5, pp. 4040--4053, 2022.

\bibitem{Bjoerk22}
J.~Björk, D.~V. Pombo, and K.~H. Johansson, ``Variable-speed wind turbine control designed for coordinated fast frequency reserves,'' \emph{IEEE Transactions on Power Systems}, vol.~37, no.~2, pp. 1471--1481, 2022.

\bibitem{Bjoerk23}
J.~Björk, K.~H. Johansson, and F.~Dörfler, ``Dynamic virtual power plant design for fast frequency reserves: Coordinating hydro and wind,'' \emph{IEEE Transactions on Control of Network Systems}, vol.~10, no.~3, pp. 1266--1278, 2023.

\bibitem{SontagWang.1995}
E.~D. Sontag and Y.~Wang, ``{On characterizations of the input-to-state stability property},'' \emph{{Systems and Control Letters}}, vol.~24, pp. 351--359, 1995.

\bibitem{SontagWang.1999}
------, ``{Notions of input to output stability},'' \emph{{Systems and Control Letters}}, vol.~38, pp. 235--248, 1999.

\bibitem{SchulteKusche+2023+891+908}
\BIBentryALTinterwordspacing
H.~Schulte and S.~Kusche, ``Fast power tracking control of pv power plants for frequency support,'' \emph{at - Automatisierungstechnik}, vol.~71, no.~10, pp. 891--908, 2023. [Online]. Available: \url{https://doi.org/10.1515/auto-2023-0029}
\BIBentrySTDinterwordspacing

\bibitem{TanakaWang}
K.~Tanaka and H.~Wang, \emph{Fuzzy Control Systems Design and Analysis: A Linear Matrix Inequality Approach}, 01 2001.

\bibitem{NREL5MW}
\BIBentryALTinterwordspacing
J.~Jonkman, S.~Butterfield, W.~Musial, and G.~Scott, ``Definition of a 5-mw reference wind turbine for offshore system development,'' 2 2009. [Online]. Available: \url{https://www.osti.gov/biblio/947422}
\BIBentrySTDinterwordspacing

\bibitem{FLEX}
A.~D. Wright, ``Modern control design for flexible wind turbines,'' \emph{National Renewable Energy Laboratory}, 2004.

\bibitem{Poeschke2022}
F.~P{\"o}schke, ``Takagi-sugeno methods with application to wind power systems,'' doctoralthesis, Leuphana Universit{\"a}t L{\"u}neburg, Universit{\"a}tsbibliothek der Leuphana Universit{\"a}t L{\"u}neburg, 2022.

\bibitem{SeDuMi}
J.~F. Sturm, ``Using sedumi 1.02, a matlab toolbox for optimization over symmetric cones.'' \emph{Optimization Methods and Software}, 1999.

\bibitem{PoeschkeGauterinSchulte2019}
F.~P\"oschke, E.~Gauterin, and H.~Schulte, \emph{New Trends in Observer-based Control}.\hskip 1em plus 0.5em minus 0.4em\relax Academic Press, August 2019, ch. LMI Region-based non-linear disturbance observer with application to robust wind turbine control, pp. 35--75.

\bibitem{DIN_EN_IEC_61400-1}
{DIN EN 61400-1:2019}, ``Wind energy generation systems - part 1: Design requirements ({IEC} 61400-1:2019).''

\bibitem{NCFG}
\BIBentryALTinterwordspacing
{COMMISSION REGULATION (EU) 2016/631}, ``establishing a network code on requirements for grid connection of generators,'' April 2016. [Online]. Available: \url{https://eur-lex.europa.eu/legal-content/EN/TXT/?uri=CELEX%3A32016R0631}
\BIBentrySTDinterwordspacing

\bibitem{FRT2018}
A.~A. MERI, Y.~AMARA, and C.~NICHITA, ``Impact of fault ride-through on wind turbines systems design,'' in \emph{2018 7th International Conference on Renewable Energy Research and Applications (ICRERA)}, 2018, pp. 567--575.

\bibitem{Poeschke22}
\BIBentryALTinterwordspacing
F.~P\"oschke and H.~Schulte, ``Evaluation of different power tracking operating strategies considering turbine loading and power dynamics,'' \emph{Wind Energy Science}, vol.~7, no.~4, pp. 1593--1604, 2022. [Online]. Available: \url{https://wes.copernicus.org/articles/7/1593/2022/}
\BIBentrySTDinterwordspacing

\bibitem{ichalal2010}
\BIBentryALTinterwordspacing
D.~Ichalal, B.~Marx, J.~Ragot, and D.~Maquin, ``{State estimation of Takagi--Sugeno systems with unmeasurable premise variables},'' \emph{{IET Control Theory and Applications}}, vol.~4, no.~5, pp. 897--908, May 2010. [Online]. Available: \url{https://hal.science/hal-00482061}
\BIBentrySTDinterwordspacing

\bibitem{gauterin2016}
\BIBentryALTinterwordspacing
E.~Gauterin, P.~Kammerer, M.~Kühn, and H.~Schulte, ``Effective wind speed estimation: Comparison between kalman filter and takagi–sugeno observer techniques,'' \emph{ISA Transactions}, vol.~62, pp. 60--72, 2016, sI: Control of Renewable Energy Systems. [Online]. Available: \url{https://www.sciencedirect.com/science/article/pii/S001905781500292X}
\BIBentrySTDinterwordspacing

\end{thebibliography}

\end{document}